\documentclass[11pt,a4paper]{article}
\usepackage[utf8]{inputenc}
\usepackage{amsmath}
\usepackage{placeins}
\usepackage{url}
\usepackage{natbib}
\usepackage{color,xcolor,setspace,geometry}
\setcitestyle{authoryear,open={(},close={)}}
\usepackage{cite}
\usepackage{amsfonts}
\usepackage[normalem]{ulem}
\usepackage{pifont}

\usepackage{arydshln}
\usepackage{multirow}
\usepackage{booktabs}

\usepackage{tikz}
\usetikzlibrary{shapes}
\usetikzlibrary{plotmarks}
\usetikzlibrary{tikzmark,matrix,calc}
\usetikzlibrary{arrows.meta,calc,decorations.markings,math,arrows.meta}

\usepackage{amssymb}
\usepackage{multirow}
\usepackage{graphicx}
\usepackage{soul}
\soulregister\citep7
\soulregister\cite7
\soulregister\citet7
\soulregister\citealp7
\soulregister\ref7
\usepackage{rotating}
\usepackage{setspace}
\usepackage{threeparttable}
\usepackage{authblk}
\usepackage{subfigure}

\usepackage{appendix}
\usepackage{bm}

\begin{document}
\begin{spacing}{1.5}	

\title{Measuring football fever through wearable technology:\\ A case study on the German cup final}

\author[1,+]{Timo Adam}
\author[1,+]{Jonas Bauer}
\author[2,*,+]{Christian Deutscher}
\author[1,3,+]{Christiane Fuchs}
\author[1,+]{Tamara Schamberger}
\author[1,2,+]{David Winkelmann}

\affil[1]{\small
Bielefeld University, Faculty of Business Administration and Economics, Bielefeld, 33615, Germany}
\affil[2]{Bielefeld University, Faculty of Sports Science, Bielefeld, 33615, Germany}
\affil[3]{Institute of Computational Biology, Helmholtz Zentrum München, Neuherberg, Germany}

\affil[*]{Corresponding author: christian.deutscher@uni-bielefeld.de}

\affil[+]{All authors contributed equally to this work and have been ordered alphabetically.}
\maketitle
\noindent
\normalsize

\begin{abstract}
Football is the world's most popular sport, evoking strong physiological and emotional responses among its fans. Yet, the specific dynamics of fan attachment to matches have received little attention in the literature. In this paper, we quantify these dynamics through a unique case study from professional football: the 2025 cup final of the German Football Association (DFB) between first-division club VfB Stuttgart and third-division club Arminia Bielefeld. We collected high-resolution smartwatch data, including heart rate and stress level, from 229 Arminia Bielefeld fans over approximately 12 weeks, complemented by survey responses on club attachment, match attendance, and personal characteristics from a subset of 37 participants. By combining physiological data with survey information, we analyse variations in emotional engagement across individuals and contexts, as well as physiological reactions to key match events. This approach provides rare, data-driven insights into the \textit{football fever} that captivates fans during high-stakes competitions. Furthermore, we compare the vital parameters recorded on the day of the match with baseline levels on non-matchdays throughout the entire observation period. Our findings reveal pronounced physiological responses among fans, beginning hours before the match and peaking at kick-off.

\end{abstract}

\section*{Introduction}

For fans, football is an emotional ride --- full of joy, frustration, hope, and pride --- whether they are cheering in the stands, watching from home, or keeping up with updates on their phones. This collective enthusiasm fuels a thriving football economy, from ticket sales and merchandise to media rights and sponsorship deals \citep{sauer2024creating}. Unsurprisingly, the cultural and commercial significance has inspired a growing body of research that explores football as both an emotional experience and a social phenomenon \citep{otting2023copula, michels2023bettors}. This paper seeks to quantify physiological and emotional responses through a unique case study in German professional football, examining their intensity, variability, and underlying drivers. It draws on data from fan engagement during the 2025 cup final in Germany, which was arguably the most important match in the history of Arminia Bielefeld, one of the two participating clubs.

The cup of the German Football Association (\textit{DFB-Pokal}) is a prestigious national competition featuring 64 teams. The tournament follows a single-elimination format and includes all clubs from the first two national divisions, along with the winners of regional cup competitions. Traditionally, the cup final is staged at the end of each season at Berlin's \textit{Olympiastadion}, the largest stadium in Germany's capital, and constitutes a nationwide spectacle watched by millions of fans. On May 24, 2025, VfB Stuttgart, the runner-up of the previous national league season and strong favourite, faced off against third-division club Arminia Bielefeld. This marked only the fourth occasion since the competition's inception in 1935 that a third-division club reached the final. For Arminia Bielefeld, their first appearance in the cup final was an extraordinary achievement that captured national attention and enthused the entire region surrounding Bielefeld for weeks.

The literature documents that fans often develop strong psychological bonds with their team, with concepts such as team identification and identity fusion describing how fans feel ``at one'' with their club \citep{newson2020devoted}. This deep attachment amplifies emotional reactions to match events, triggering substantial physiological responses. Research has recorded increases in spectators’ heart rate, blood pressure, and stress hormones during high-stakes matches \citep{van2012testosterone}. Fans who are firmly fused with their team experience the highest release of stress hormones when under game stress \citep{newson2020devoted}. In extreme cases, the cardiovascular strain of spectating can be severe, leading to an increased incidence of cardiac emergencies \citep{maturana2021hard, mahevcic2022incidence}. Notably, the communal rituals surrounding matches can evoke powerful shared emotions: For example, the excitement of Brazilian football fans (measured via heart rate arousal) has been shown to not peak during the match itself but during pre-match fan rituals, with only goal celebrations matching that pre-match high \citep{xygalatas2025route}. Summarising, this literature underscores how profoundly sports fandom engages both mind and body, providing a foundation for quantifying fans’ physical and emotional responses in our case study.

This paper examines the emotional attachment of sports fans to events involving their favourite club. Unlike the studies discussed above, it utilises smartwatch data. This allows us to track over 200 fans, many more than the typical sample sizes of up to 50 participants in previous research. In addition, the tracking data spans several weeks, not just the day of the match, enabling us to isolate the effects of the match from other dynamics. The 2025 cup final serves as a natural experiment in our study, given its exceptional importance to Arminia Bielefeld fans. In collaboration with the \textit{Wissenswerkstadt Bielefeld}, we recorded vital parameters from Arminia Bielefeld fans over a 12-week period before, during, and after the cup final using Garmin smartwatches. Additionally, participants were surveyed to assess their attachment to Arminia Bielefeld, attendance at the cup final, and personal characteristics. We received responses to the questionnaire from a subset of participants, allowing us to link survey information with the smartwatch data. This linkage enables us to draw conclusions about the emotional attachment of sports fans based on their characteristics and factors such as the location where they watched the match.

\section*{Data collection}
\label{sec:data}

Our study aims to explore the emotional attachment of Arminia Bielefeld fans to the cup final. To this end, fans who intended to watch the match were recruited and invited to share their physiological data with us. After the final, participants were asked to complete a survey and provide additional information to enable a more in-depth analysis. This section details the procedure for collecting the smartwatch data and outlines the survey distributed to participants, along with summarised responses.

\subsection*{Smartwatch data}

Our study comprises 229 Garmin smartwatch users of legal age who declared themselves as fans of Arminia Bielefeld. Participants were recruited via local and national media reports. To contextualise the findings from the cup final with respect to vital parameters observed on \textit{regular} days, data were collected over an extensive period, commencing on May 14, 2025, 10 days before the cup final, and concluding on July 31, 2025. Here and in the following, a \textit{regular} day refers to days without an official match of Arminia Bielefeld and without local holidays. The prolonged observation period enables us to analyse the day of the cup final in relation to general physiological and behavioural patterns. Participants consented to provide their data for the duration of the study by synchronising their smartwatch with the corresponding smartphone application. The number of active participants slightly varies over time, as not all fans contributed data throughout the entire observation period; specifically, we received data from 194 participants during the cup final itself. The study was conducted by Bielefeld University in accordance with relevant institutional, national, and international guidelines and regulations. Before data collection, informed consent was obtained from all participants, and informed consent was obtained from the participants to publish the analysis of anonymised data.

The smartwatch records various sports-related metrics, including active seconds, motion intensity, and hourly steps. For this study, we primarily focus on the heart rate and stress level as indicators of fans' emotional attachment to the match. This choice is consistent with previous research demonstrating the suitability of these variables for capturing emotional responses \citep{appelhans2006heart}. The heart rate is reported as beats per minute (bpm) and recorded in 15-second intervals, whereas the stress level is recorded in three-minute intervals. The stress level tracked by the smartwatch is estimated from heart rate variability and expressed on a scale from 0 to 100, with 0 representing no stress and 100 indicating maximum stress. Notably, the stress level is not recorded when the person wearing the smartwatch is highly active (e.g.\ during exercise), as physical exertion itself induces substantial heart rate variability. Descriptive statistics on both stress levels and heart rates are provided in Table~\ref{tab:sum_stats} and Figure~\ref{fig:stress_week} below.

\subsection*{Survey of participants}

The measured vital parameters of fans are likely influenced by individual characteristics, fan- and match-related factors. To capture these variables, we distributed a survey (full questionnaire provided in the supplementary material) to 95 study participants who had consented to further contact after an updated data privacy statement was provided. The survey is structured into three main categories: (1) personal characteristics, (2) match-related variables, and (3) cup final-specific questions.

Given the established link between age and heart rate \citep{tanaka2001age,reardon1996changes}, we incorporated personal characteristics within category (1) of the survey. Beyond age and gender, we expect the degree of engagement with Arminia Bielefeld to influence fans' physiological responses during the match. Research in basketball, for instance, has shown that in-person spectatorship produces greater group physiological synchrony and more intense ``transformative'' experiences than viewing in a small remote group \citep{baranowski2022being}. In our case, a dedicated fan who regularly attends matches of Arminia Bielefeld in the stadium might exhibit a different physiological reaction to match dynamics compared with someone who rarely attends in person. We therefore asked for the attachment to Arminia Bielefeld, operationalised through club membership status and the number of matches attended in person during the 2024/25 season. 

Under category (2) of the survey, we asked participants where they had followed the cup final. Those who reported watching the match (either on TV, at a public gathering, or in the stadium) were also asked whether they had consumed alcohol \citep{struven2025impact}. Category (3) applied exclusively to respondents who attended the final in the stadium. These participants were asked about their arrival time in the host city to capture potential travel-related impact on the vital parameters. Media reports indicated that overcrowding at stadium entrances caused delays, potentially evoking feelings of danger, anxiety, or frustration, which could in turn influence physiological parameters. Accordingly, we asked participants when they joined the entrance queue and how long the entry process lasted. In addition, participants were asked about betting behaviour related to the cup final; however, no one reported having placed a bet.

The survey was distributed to 95 study participants, of whom 37 responded within the requested 16 days: 17 female and 20 male, aged 18 to 63 (average age 38.7 years). Of all participants, 33 were employed, with 30 of them starting work between 6 and 9~a.m. The end of a typical workday for these 30 participants exhibits a much higher variation, occurring between 12:30 and 7~p.m. However, 20 participants indicated that their usual workday ended between 4 and 6~p.m. Among the respondents, 18 (48.6\%) were Arminia Bielefeld members and 14 (37.8\%) were season ticket holders; 12 participants were both.

Except for one survey participant, all of them watched the cup final: 20 attended the match in the stadium, five joined public gatherings, and 11 watched the match on TV. 50\% reported consuming alcohol during the cup final, with a notably higher rate of 65\% among those attending in the stadium. Furthermore, 77.8\% of club members and 78.6\% of season ticket holders attended the match in the stadium. Additionally, all 11 fans who attended more than 10 matches of Arminia Bielefeld in stadiums during the 2024/25 season also watched the cup final in the stadium in Berlin in person, compared with less than 50\% of those who attended only 1--10 matches during the season. The cup final took place on a Saturday evening, starting at 8~p.m. Of the 20 participants who attended the match in person, nine arrived in Berlin on Friday or earlier, while 11 arrived on the matchday. Eighteen fans participated in the fan festival before the match, which was held in a central location in Berlin.

\section*{Results: Fans' attachment to the cup final}
Collecting data from smartwatches worn by study participants enables us to investigate their attachment to the 2025 cup final by analysing stress levels and heart rates during the match and comparing these with patterns observed on regular days. In this section, we (1) present summary statistics on participants' stress levels and heart rates, (2) investigate their patterns on the day of the cup final compared with the entire observation period to interpret findings in a broader context, and (3) analyse vital parameters in more detail over the course of the cup final itself to conclude on the impact of the match progression.

\subsection*{Summary statistics}

\begin{table}[ht]
    \centering
    \hspace*{-0.05\textwidth}
    \scalebox{0.85}{
    \begin{tabular}{llrrrrrr}
    \toprule
      {\textbf{Variable}} & \textbf{Time window} & \textbf{Min.} & \textbf{1. Quartile} & \textbf{Median} & \textbf{Average} & \textbf{3. Quartile} & \textbf{Max.} \\
      \midrule
    \multirow{3}*{Stress level} & Cup final &  0.0 & 21.0 & 40.0 & 44.2 & 67.0 & 100.0\\
    & Regular day &0.0 & 16.0 & 24.0 & 31.1 & 43.0 & 100.0 \\
         & Overall & 0.0 & 16.0 & 25.0 & 31.3 & 43.0 & 100.0\\
         \midrule
         \multirow{3}*{Heart rate} & Cup final & 35.0  &  64.0 &   77.0 &   78.7 &   91.0 &  181.0 \\
        & Regular day & 27.0&   58.0&   68.0&   70.9&   80.0 & 243.0 \\
         & Overall &27.0&      58.0&      69.0&      71.0&      81.0&     243.0 \\
        \bottomrule
    \end{tabular}}
    \caption{Summary statistics of stress levels (scale from 0 to 100) and heart rates (bpm) for individual participants, measured at single time points, on the day of the cup final (May 24, 2025) and on regular days within the observation period (May 14 to July 31, 2025).}
    \label{tab:sum_stats}
\end{table}


Table~\ref{tab:sum_stats} presents summary statistics of the individual stress levels and heart rates. One observation represents the corresponding value of one participant at a single time point. We divide the data into the day of the cup final and regular days. All descriptive statistics for stress levels are higher on the day of the cup final, with the average stress level (45.3) being increased by approximately 42\% compared with regular days (31.9) and close to the maximum observed on any regular day (48.2). Similarly, heart rate statistics are generally elevated on the day of the cup final, with the exception of the maximum value.

\begin{figure}[htp!]
    \centering
    \includegraphics[width=0.5\linewidth]{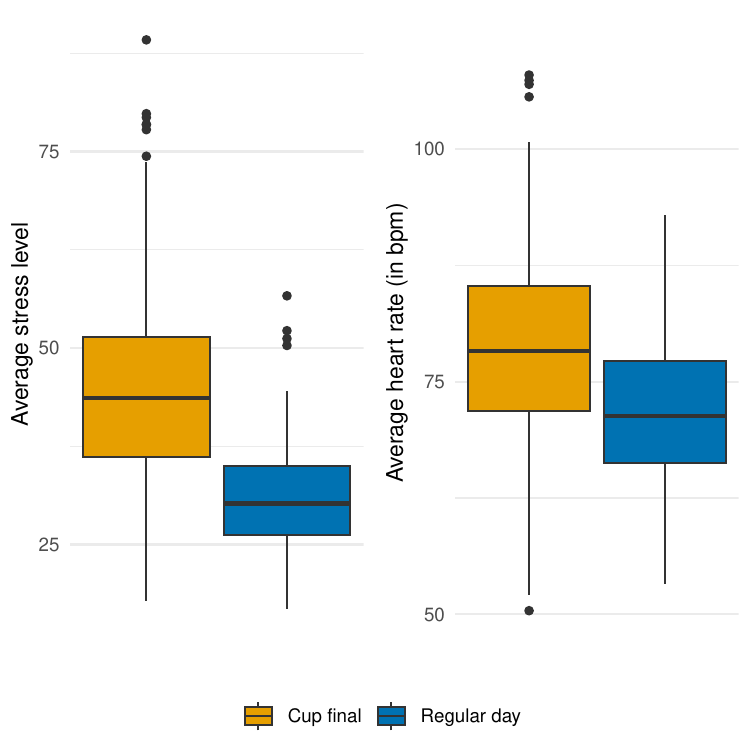}
    \caption{Boxplot of the average stress level (left) and average heart rate (right) per individual participant across time points for the day of the cup final (May 24, 2025; orange) and regular days within the observation period (May 19 to July 31, 2025; blue).}
    \label{fig:boxplot_individual}
\end{figure}

Figure~\ref{fig:boxplot_individual} illustrates the distribution of the average stress levels (left panel) and average heart rates (right panel) for each participant across time points using box plots comparing the day of the cup final to regular days. We find both the average heart rate and average stress level to be elevated on the day of the cup final. While the maximum average stress level on regular days is approximately 50, values rise to up to 90 for single individuals on the day of the cup final. Notably, the variation in stress levels among study participants is much greater on the day of the cup final, potentially influenced by the different locations where they attended the match. For the relation between heart rates and stress levels on the day of the cup final and regular days, see Figure~S1 in the Supplement.

\subsection*{Vital parameters over the course of the matchday in light of regular days}
\label{sec:comparison}

\begin{sidewaysfigure}
     \centering
     \includegraphics[width=1\linewidth]{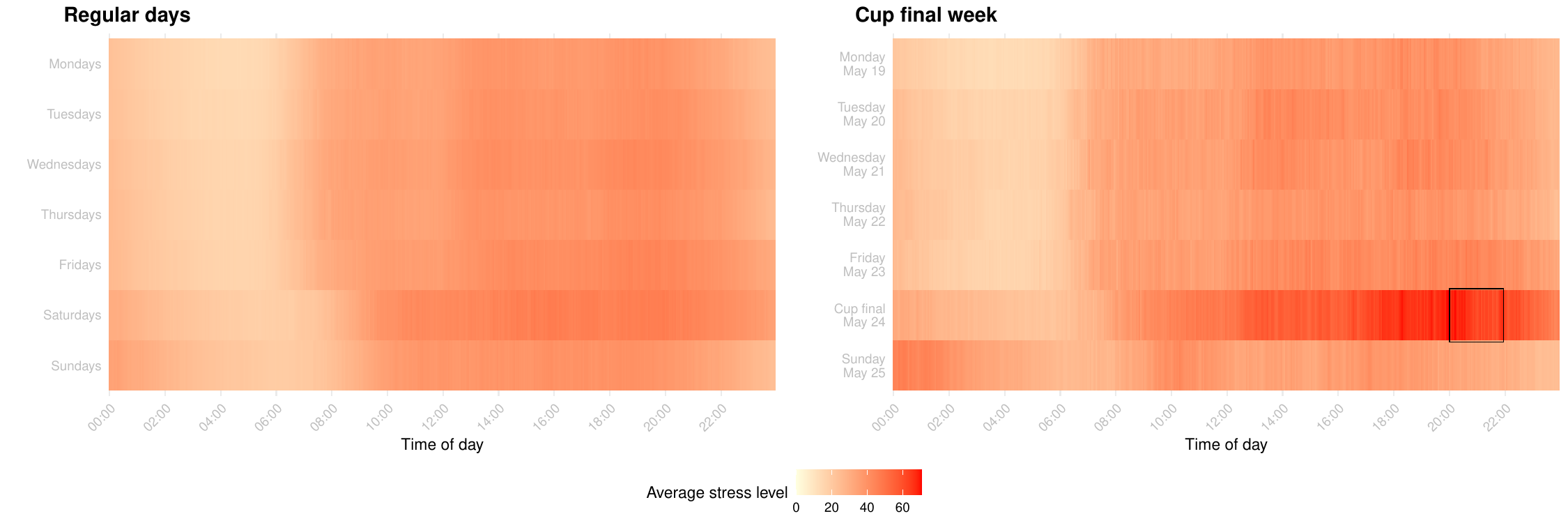}
     \caption{Average stress levels across participants and across regular days between May 19 and July 31, 2025 (left panel) and average stress levels across participants over the course of the day from May 19 to May 25, 2025 (right panel). The black box highlights the cup final on May 24.}
     \label{fig:stress_week}
 \end{sidewaysfigure}

In this section, we compare vital parameters on the cup final matchday with those on regular days within the observation period, thereby providing a basis for understanding how matchdays alter typical physiological dynamics. Figure~\ref{fig:stress_week} illustrates the average stress levels across participants and across regular days throughout the study's observation period (left panel) and average stress levels across participants during the week of the cup final (right panel). Darker colours indicate higher average stress levels. As we observe variation in participants' stress level patterns depending on the day of the week, we average stress levels in the left panel not only across all participants but also across the same weekdays of different weeks. Consequently, the values for regular days represent the average stress levels across all participants over the 12 weeks of data collection and across the respective day of the week. Averages per time point are calculated based on the number of applicable measurements, which vary slightly. 

Recurring patterns are evident across weekdays, with stress levels generally lowest at night, as expected when most people are asleep. Additionally, we observe clear differences between typical weekdays (Monday to Friday) and weekends (Saturday and Sunday). On weekdays, average stress levels start to increase at around 6~a.m., whereas on weekends, they do not rise before 8~a.m., likely due to people sleeping in. Furthermore, average stress levels on regular Saturday nights are noticeably higher than on weeknights, probably because people are more likely to go out. Overall, Saturdays appear to be the most stressful days, showing higher average stress levels than other days of the week during time awake, even in weeks without a football match. This pattern may be explained by greater activity on Saturdays. In contrast, weekdays tend to be less stressful, with average stress levels on Sundays comparable to those observed on weekdays.

We now compare the week of the cup final (right panel) with averages across regular days (left panel). Overall, the two panels show broadly similar patterns for most of the time. However, the day of the cup final on May 24 stands out clearly, highlighting the event's impact on participants' average stress level. The average is considerably higher on the matchday than on regular Saturdays, with the strongest increase observed in the two hours before kick-off and during the match itself. Notably, an elevated average stress level is already visible during the preceding night, which may be partly driven by some fans travelling from Bielefeld to Berlin the day before and therefore staying awake for longer than usual. Moreover, on the day of the cup final, the average stress level is higher than on any regular day, even starting from typical wake-up times. While this early increase may partly be attributed to some fans travelling to Berlin, it further suggests that fans are already excited about the match several hours before kick-off. The average stress level continues to rise over the course of the day, particularly after lunch, and peaks between 6 and 8~p.m., right before kick-off. This peak likely reflects both heightened tension and increased fan mobility, such as going to the stadium, public gatherings, friends' houses, or welcoming guests. The average stress level remains elevated during and after the match, exceeding averages at any point on regular days. This sustained elevation is plausibly linked to the intense emotions associated with the match experience.

\begin{figure}[htp!]
    \centering
    \includegraphics[width=0.9\linewidth]{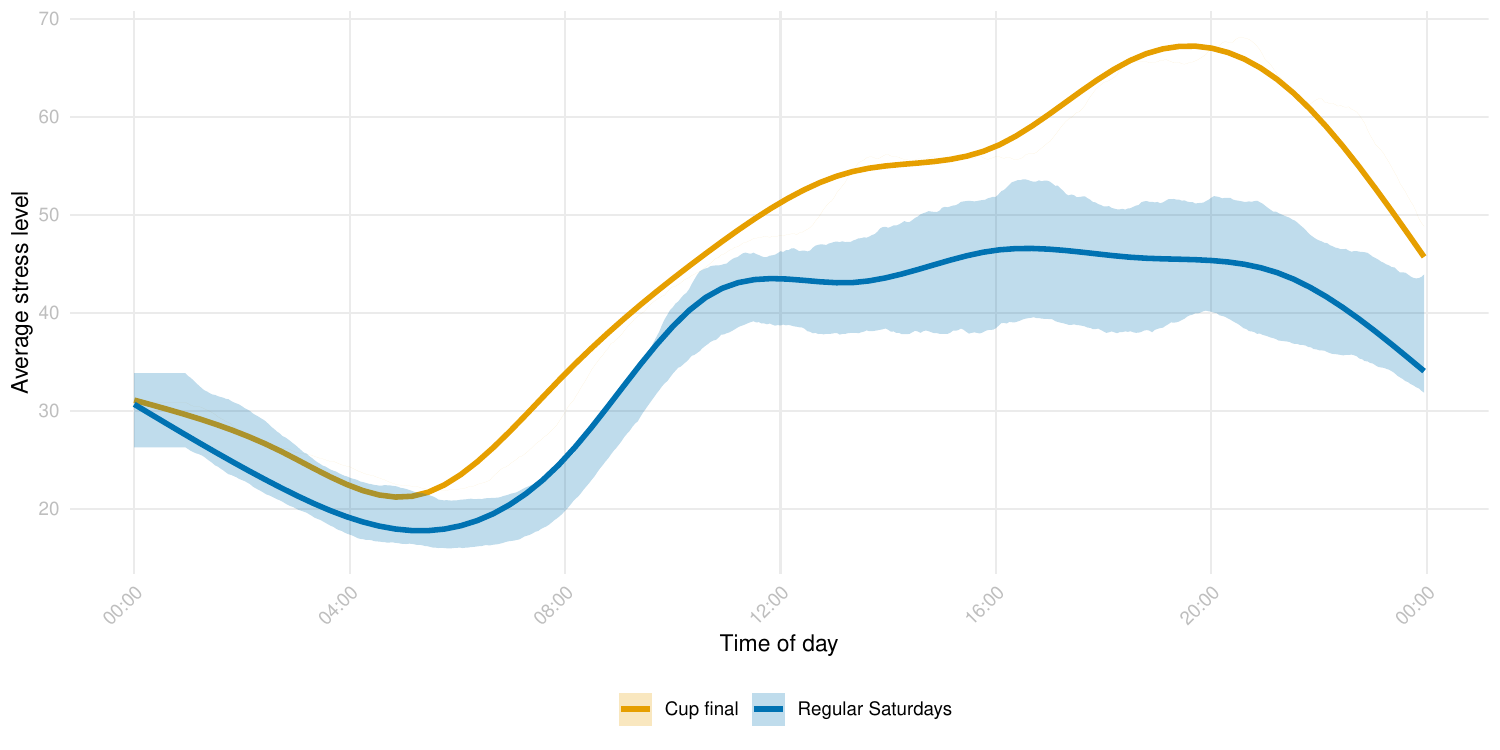}
    \caption{Average stress level across participants over the course of the cup final matchday (May 24, 2025; orange) compared with the average stress level across participants and across all regular Saturdays within the observation period (May 19 to July 31, 2025; blue). The shaded areas represent the 10th- and 90th-percentiles of averages across participants for regular Saturdays. For the graphical representation, we smooth averages and percentiles to reduce volatility and increase interpretability.}
    \label{fig:stress_saturdays}
\end{figure}

Next, we evaluate the difference between the Saturday on which the cup final took place and regular Saturdays --- the days that are, on average, the most stressful during the week anyway --- in more detail. Figure~\ref{fig:stress_week} illustrates the average stress levels across participants on the day of the cup final compared with the average stress levels across participants on regular Saturdays, with the latter being averaged across both individuals and multiple Saturdays. In addition, it displays the 10th- and 90th-percentiles across Saturdays: for each Saturday and each time point of the day, stress levels were first averaged across participants, yielding one value per time point and Saturday; the percentiles were then computed across these values at each time point. Similar to our findings in Figure~\ref{fig:stress_week}, we observe higher average stress levels on the day of the cup final than on regular Saturdays at every time of the day. This difference is particularly pronounced before kick-off and during the match. Specifically, during the match, i.e.\ between 8 and 10~p.m., the average stress level is higher by 43\% compared with regular Saturdays. Moreover, during the typical time awake, the average stress levels on the day of the cup final exceed the 90th-percentile of regular Saturdays at every time point. Consequently, the day of the cup final appears to be more stressful for participants than other Saturdays, most likely due to the emotions surrounding their club's participation in this outstanding event.

\subsection*{Vital parameters over the matchday and the course of the match}
\label{sec:cupfinal}

The final of the German cup competition on May 24, 2025, attracted considerable national and international attention, primarily due to the participation of third-division club Arminia Bielefeld. Given the exceptional significance of this event for Arminia Bielefeld fans, this section takes a closer look at the cup final itself and the responses of supporters attending the match. We first outline the course of the cup final, then analyse fans' vital parameters over the match period in relation to the uncertainty of the match outcome and across different viewing contexts, and finally examine whether the day of arrival in the host city influences fans' stress levels for those attending in the stadium.

According to pre-match betting odds, provided by a major European bookmaker, VfB Stuttgart, the runner-up of the previous national league season and opponent of Arminia Bielefeld, was deemed the clear favourite. Specifically, translating these betting odds into odds-implied winning probabilities as a predictor for the match outcome \citep{winkelmann2024betting}, the estimated probability of a VfB Stuttgart victory was 70\%, a draw after 90 minutes was about 20\%, and an Arminia Bielefeld win was approximately 10\%. Although the initial phase of the match was quite balanced, with even promising scoring opportunities for Arminia Bielefeld, VfB Stuttgart scored in the 14th, 22nd, and 27th minutes. This 3-0 lead increased VfB Stuttgart's odds-implied winning probability to over 95\%, rising to more than 97\% by halftime. In the second half, VfB Stuttgart scored again in the 67th minute, before Arminia Bielefeld was able to reduce the deficit with goals in the 84th and 87th minutes, eventually resulting in a 4–2 victory for VfB Stuttgart. Despite these being the first goals ever scored by a third-division club in the cup final, they had minimal impact on the odds-implied probabilities: Prior to Arminia Bielefeld's first goal, the probabilities were 99.5\% for a VfB Stuttgart win and 0.25\% each for a draw and an Arminia Bielefeld victory. Following the 4–2 scoreline, VfB Stuttgart's winning probability only slightly decreased to 97.7\%, with a 2.0\% probability of a draw and 0.3\% for an Arminia Bielefeld win.

\begin{figure}[htp!]
    \centering    
    \includegraphics[width=0.9\linewidth]{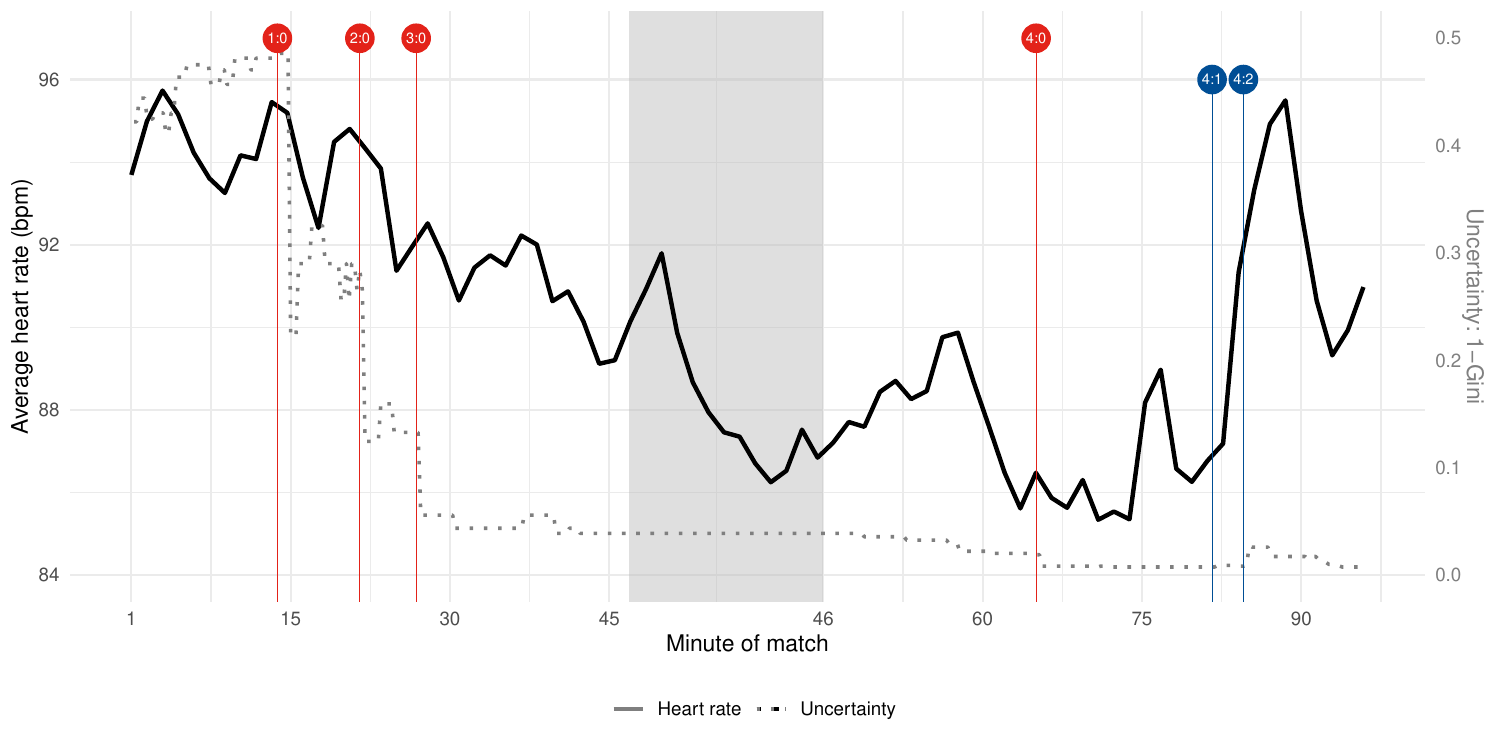}
    \caption{Average heart rate across participants over the course of the match (black solid line; left y-axis) and objective uncertainty of the match outcome derived from betting odds measured by $1-\text{Gini}$ (grey dotted line; right y-axis) during the cup final. For the graphical representation, we smooth the average heart rate to reduce volatility and increase interpretability. The grey shaded area marks the halftime break. Goals are indicated by vertical lines (red: VfB Stuttgart; blue: Arminia Bielefeld), together with the resulting scoreline.}
    \label{fig:uncertainty}
\end{figure}

Figure~\ref{fig:uncertainty} illustrates the average heart rate of those study participants who provided data throughout the cup final, starting from kick-off at 8~p.m. (minute 1) until the end of the added time (around 10~p.m.). To relate fan attachment to uncertainty about the match outcome, Figure~\ref{fig:uncertainty} additionally depicts the measure $1-\text{Gini}$ to quantify (objective) uncertainty (grey dotted line, right y-axis). The Gini coefficient is calculated based on odds-implied probabilities and ranges from 0 to 1, with higher values of $1-\text{Gini}$ indicating a higher degree of uncertainty. For example, odds-implied probabilities of one-third for each of the three possible outcomes would yield a $1-\text{Gini}$ value of one.

Results presented in Figure~\ref{fig:uncertainty} indicate that the tension of Arminia Bielefeld fans is highest during the first 15 minutes of the match, with the average heart rate reaching up to approximately 96~bpm. After each goal scored by the opposing team, VfB Stuttgart, we observe a decrease in the heart rate, with the average value falling below 90~bpm until the beginning of the halftime break. The average heart rate increases during the first few minutes of the break, possibly due to increased movement, for example, to get refreshments. When the match resumes, however, the average heart rate remains low at about 87~bpm. It might be that fans maintained hope for a comeback of Arminia Bielefeld, coming with a slight increase in tension during the first 15 minutes of the second half, with the average heart rate rising again to around 90~bpm. However, even before VfB Stuttgart scored their fourth goal in the 67th minute, the tension among Arminia Bielefeld fans had already diminished. The average heart rate drops to its lowest point of the match shortly after the goal (below 86~bpm around the 70th minute). Especially for the first half, the subjective uncertainty of the match outcome perceived by fans, as reflected in their heart rates, aligns with the development of objective uncertainty, represented by $1-\text{Gini}$: Both lines peak within the first 15 minutes of the match and then steadily decrease until halftime. While the average heart rate declines more gradually, the objective uncertainty based on odds-implied probabilities decreases in steps following each goal. This suggests that fan tension reduces with a delay following major events during football matches.

The results in Figure~\ref{fig:uncertainty} are particularly interesting for the last 15 minutes of the match, where Arminia Bielefeld scores two goals. While the objective uncertainty of the match outcome remains very low (the value of $1-\text{Gini}$ increases from~0.007 to only~0.026 after the two goals), the average heart rate increases by around 10~bpm and reaches similar values after the second goal by Arminia Bielefeld as during the first 15 minutes of the match. This suggests that there is more fan excitement than objectively quantified uncertainty about the final match outcome. Still, fans might have overestimated the impact of these two goals on it. A different contributing factor might be pride in the extraordinary event, the first scoring in a cup final by a third-division club.

\begin{figure}[htp!]
    \centering
    \includegraphics[width=0.9\linewidth]{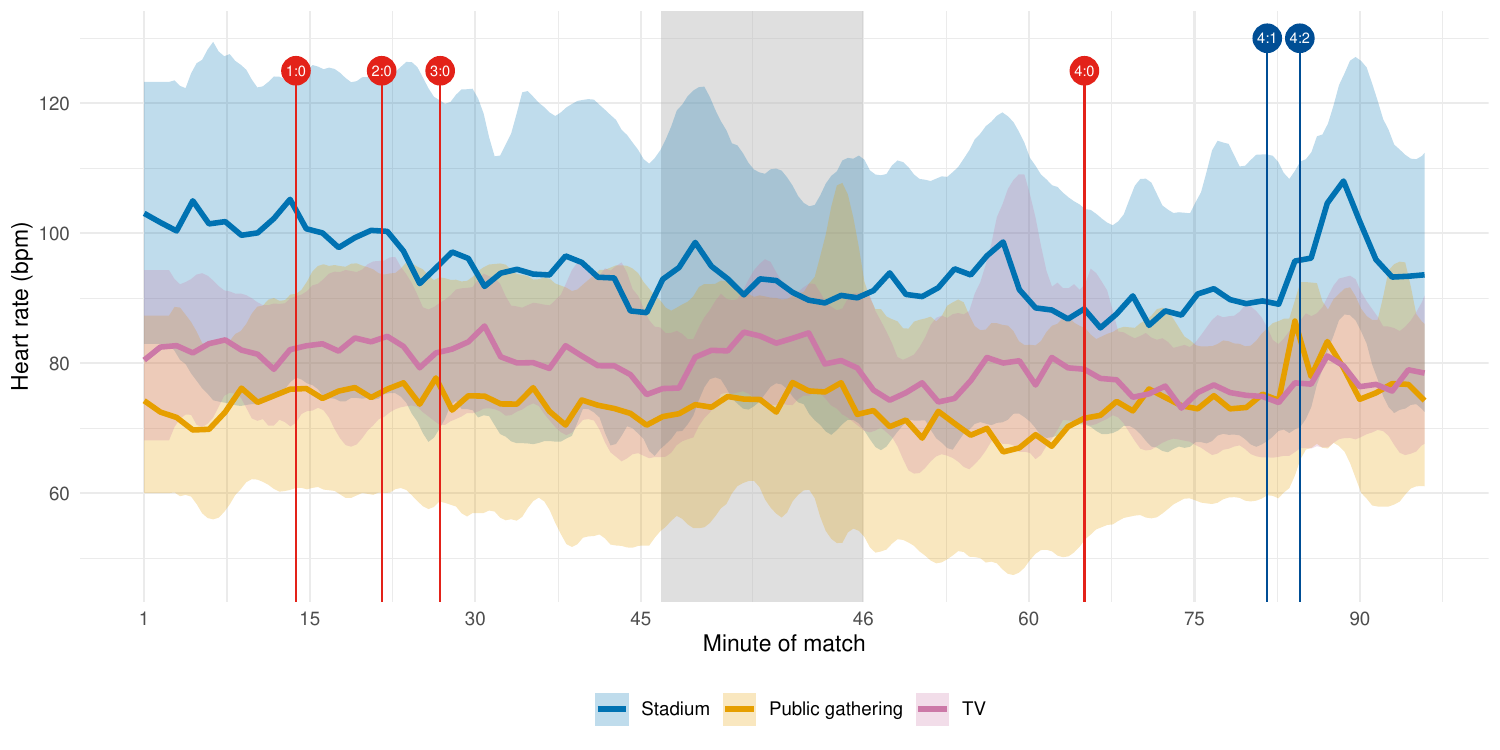}
    \caption{Average heart rate across survey participants with the same viewing context over the course of the match (blue: stadium; orange: public gathering; pink: TV). The shaded areas represent the heart rates' 10th- and 90th-percentiles. For the graphical representation, we smooth averages and percentiles to reduce volatility and increase interpretability. The grey shaded area marks the halftime break. Goals are indicated by vertical lines (red: VfB Stuttgart; blue: Arminia Bielefeld), together with the resulting scoreline.}
    \label{fig:heart_rate_by_location}
\end{figure}

To further contextualise the fan experience, Figure~\ref{fig:heart_rate_by_location} shows the average heart rate over the course of the match, while distinguishing between participants who watched the match in the stadium, at a public gathering, or on TV. These values are based on different numbers of participants, determined by the location of the match attendance, and include only those participants who filled out the additional survey. The average heart rate is highest for participants in the stadium (average: 94.2~bpm), followed by those watching TV (79.4~bpm) and at a public gathering (73.8~bpm). On average, heart rates of stadium attendees are 23.1\% higher than those of participants watching elsewhere. After the first goal of Arminia Bielefeld, this difference increases to a maximum of 35.8\%, with a maximum average heart rate of 108.0~bpm in the stadium. These findings suggest that the emotional and physiological arousal associated with watching the match is particularly pronounced in the stadium environment compared with other viewing contexts. Beyond the location where the match was watched, alcohol consumption is also associated with elevated heart rates: The average heart rate of participants who reported alcohol intake is on average 5.3\% higher throughout the match, 7.4\% higher during the second half, and even 11.7\% higher following Arminia Bielefeld's first goal. Interpretation of the heart rate should therefore take into account that alcohol intake can contribute to an acute increase in cardiovascular strain, particularly in emotionally arousing contexts: Elevated heart rates in combination with alcohol are known to increase the risk of arrhythmias and other adverse cardiac events \citep{brunner2024acute}, which may be of clinical relevance during high-stress situations such as sports events.

\begin{figure}
    \centering    
    \includegraphics[width=0.9\linewidth]{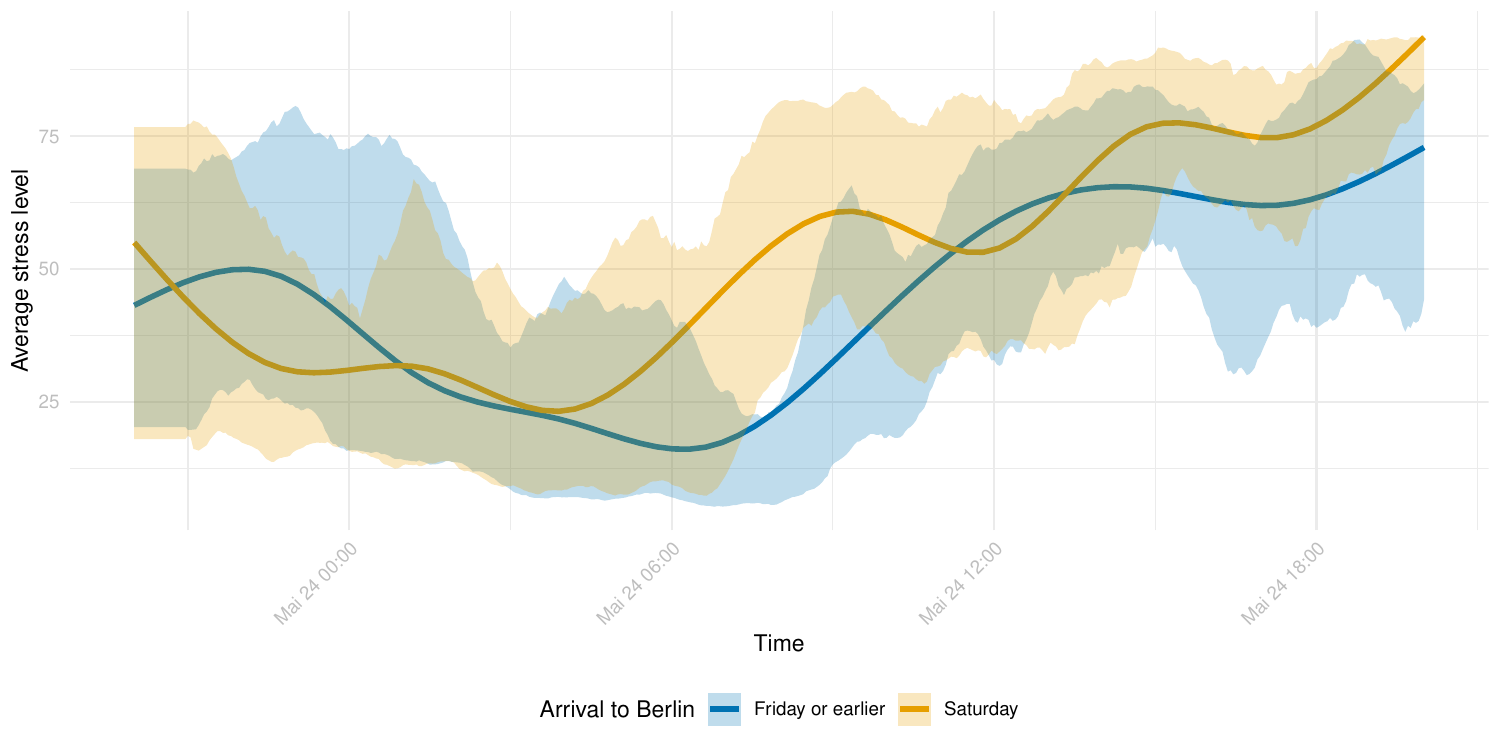}
    \caption{Average heart rates across survey participants attending the final in the stadium with the same arrival to Berlin (orange: Friday (or earlier); blue: Saturday) over the 24 hours prior to kick-off. The shaded areas represent the heart rates' 10th- and 90th-percentiles. For the graphical representation, we smooth averages and percentiles to reduce volatility and increase interpretability.}
    \label{fig:arrival}
\end{figure}

For the study participants who attended the cup final in the stadium in Berlin, we further gathered information on whether they arrived in the host city on Friday (or earlier) or on Saturday, i.e.\ the matchday. Figure~\ref{fig:arrival} illustrates the average stress level across these fans, differentiated by the day of arrival, along with the 10th- and 90th-percentiles. For both groups, the average stress level steadily increases over the course of the matchday, peaking just before kick-off. Notably, there are distinct differences depending on the day of arrival. Fans who travelled to Berlin prior to the matchday exhibit higher average stress levels late Friday evening, possibly due to travel-related efforts. In contrast, those arriving on Saturday display increased average stress levels from 6~a.m. on the matchday, while the increase of the Friday group starts around 8~a.m. only. While the average stress levels are relatively similar between 11~a.m. and 2~p.m., fans arriving on Friday or earlier exhibit lower average stress levels from 2~p.m. until kick-off. Potentially, fans staying overnight in Berlin had the opportunity to rest in their hotels. In contrast, those arriving solely on the matchday might have left the host city directly after the match and thus had to remain active the whole day, e.g.\ by visiting the fan festival of Arminia Bielefeld. In conclusion, fans travelling to the host city of the cup final only on the matchday exhibit higher average stress levels throughout the entire day.

\section*{Discussion}
\label{sec:discussion}

This study examines the emotional attachment of football fans to high-stakes matches. Drawing on the natural experiment of the 2025 German cup final, our findings demonstrate that such matches elicit pronounced physiological responses among fans. Using smartwatch data from more than 200 fans of Arminia Bielefeld, one of the two participating clubs, we observe elevated stress and heart rate levels on the matchday, clearly exceeding baseline values from \textit{regular} weekdays and weekends, i.e.\ days without an official match of Arminia Bielefeld. These increases are evident even prior to kick-off, peak during phases of the match perceived as decisive, and remain elevated well into the night after the final whistle, indicating sustained arousal throughout the entire matchday. Responses, however, vary by context: stadium attendance, alcohol consumption, and the day of arrival to the host city. Moreover, perceived uncertainty regarding the match outcome and pivotal events within the match, such as goals, further amplify the heart rate of fans. Beyond these findings, the study also demonstrates the potential of wearable technology to capture emotional engagement in large samples and over extended periods, providing continuous and non-invasive measures of physiological reactions in the context of a natural experiment.

Still, limitations remain: reliance on indirect stress measures provided by the smartwatches, a small subsample of the post-study survey, and the focus on a single exceptional event. While our study relies on smartwatch heart rate and heart rate variability-based stress measures, future research could profit from integrating broader physiological variables. Recent advances in wearable technology illustrate the potential of multimodal monitoring. For instance, research demonstrated that multispectral sensor fusion in smartwatches allows continuous assessment of hydration and sweat loss, offering new perspectives on physiological reactions during emotionally intense events\citep{volkova2023multispectral}. Similarly, hybrid wearable patches that combine electromyography with sweat cortisol sensing can capture both muscular and biochemical stress indicators\citep{hossain2024stressfit}. Incorporating such multimodal data would enable a richer and more precise characterisation of \textit{football fever}, contributing to a comprehensive understanding of fan arousal. Beyond the scope of the present work, our study opens avenues for examining vital parameters in daily life and within general natural high‑pressure contexts.

\vspace{1cm}
\noindent
\textbf{Acknowledgements:} Our research was supported by the Deutsche Forschungsgemeinschaft (DFG, German Research Foundation) -- RTG 2865 (CUDE).

\newpage
\bibliographystyle{apalike} 
\bibliography{library}

\newpage

\appendix
\begin{appendices}

\renewcommand{\thetable}{S\arabic{table}}
\renewcommand{\thefigure}{S\arabic{figure}}
\setcounter{table}{0}
\setcounter{figure}{0}

\section*{Supplementary Material}

\begin{figure}[htp!]
    \centering
    \includegraphics[width=\linewidth]{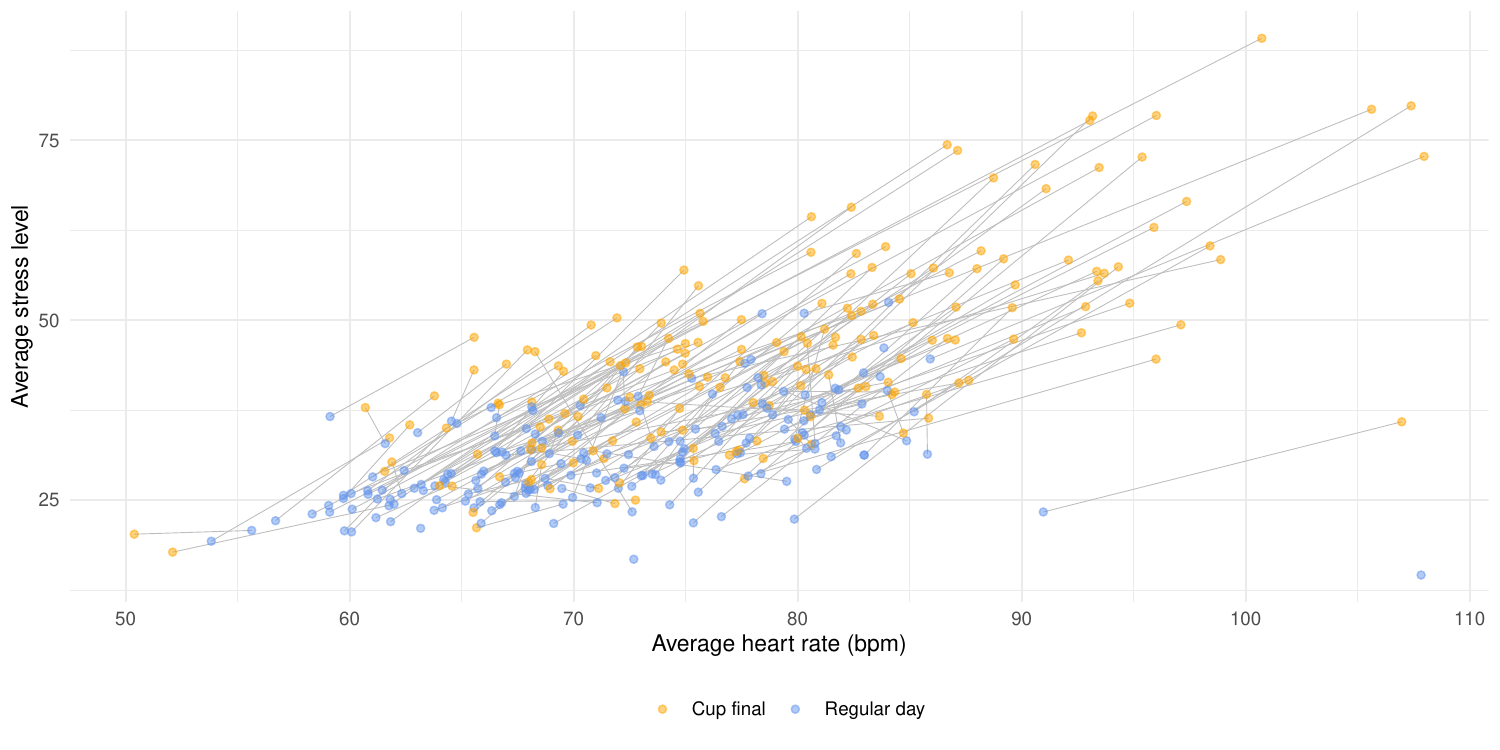}
    \caption{Relation between average heart rate and average stress level for each participant on the day of the cup final (orange) and on regular days (blue).}
    \label{fig:scatterplot}
\end{figure}

\begin{figure}
    \centering
    \includegraphics[width=\linewidth]{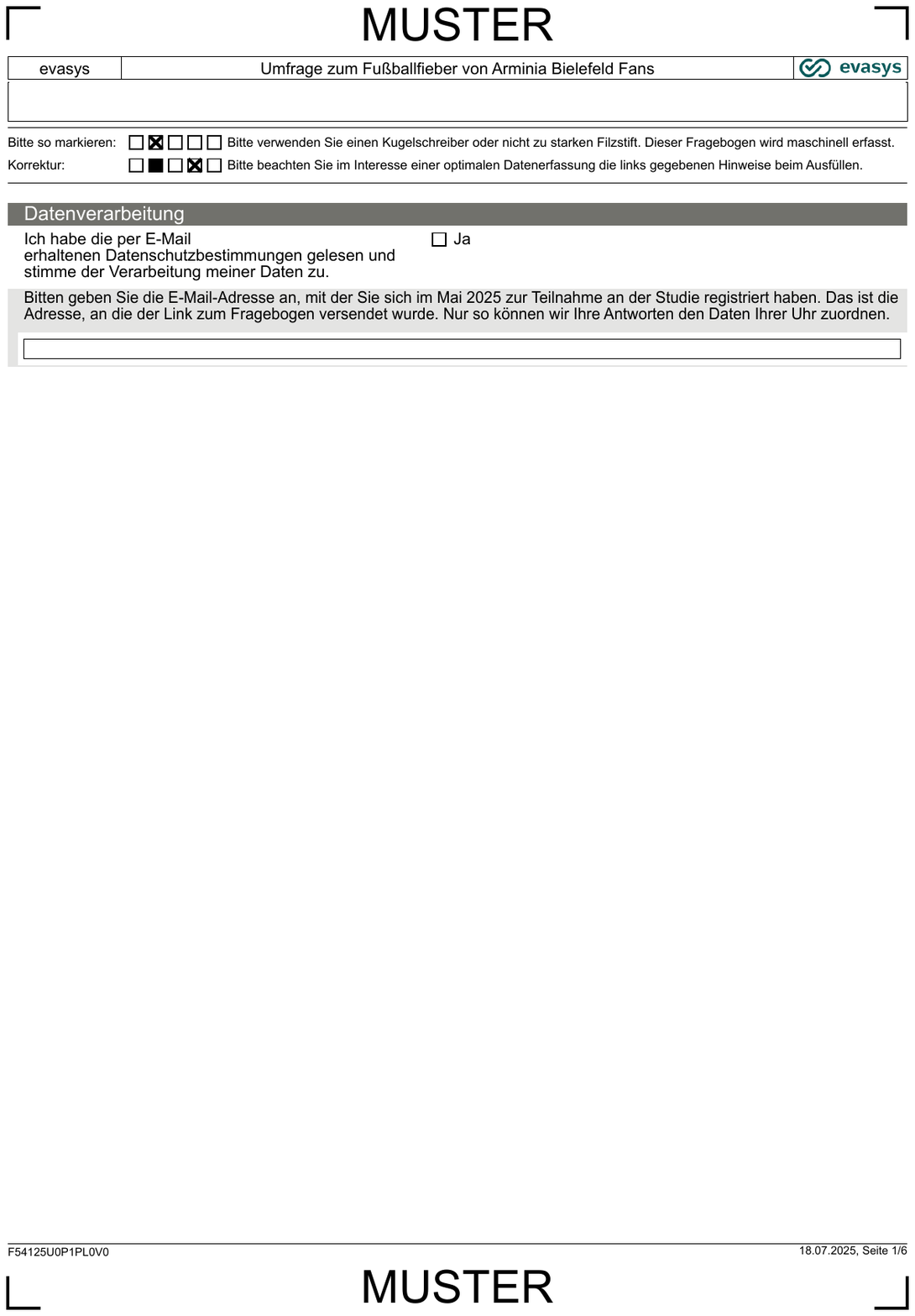}
\end{figure}

\begin{figure}
    \centering
    \includegraphics[width=\linewidth]{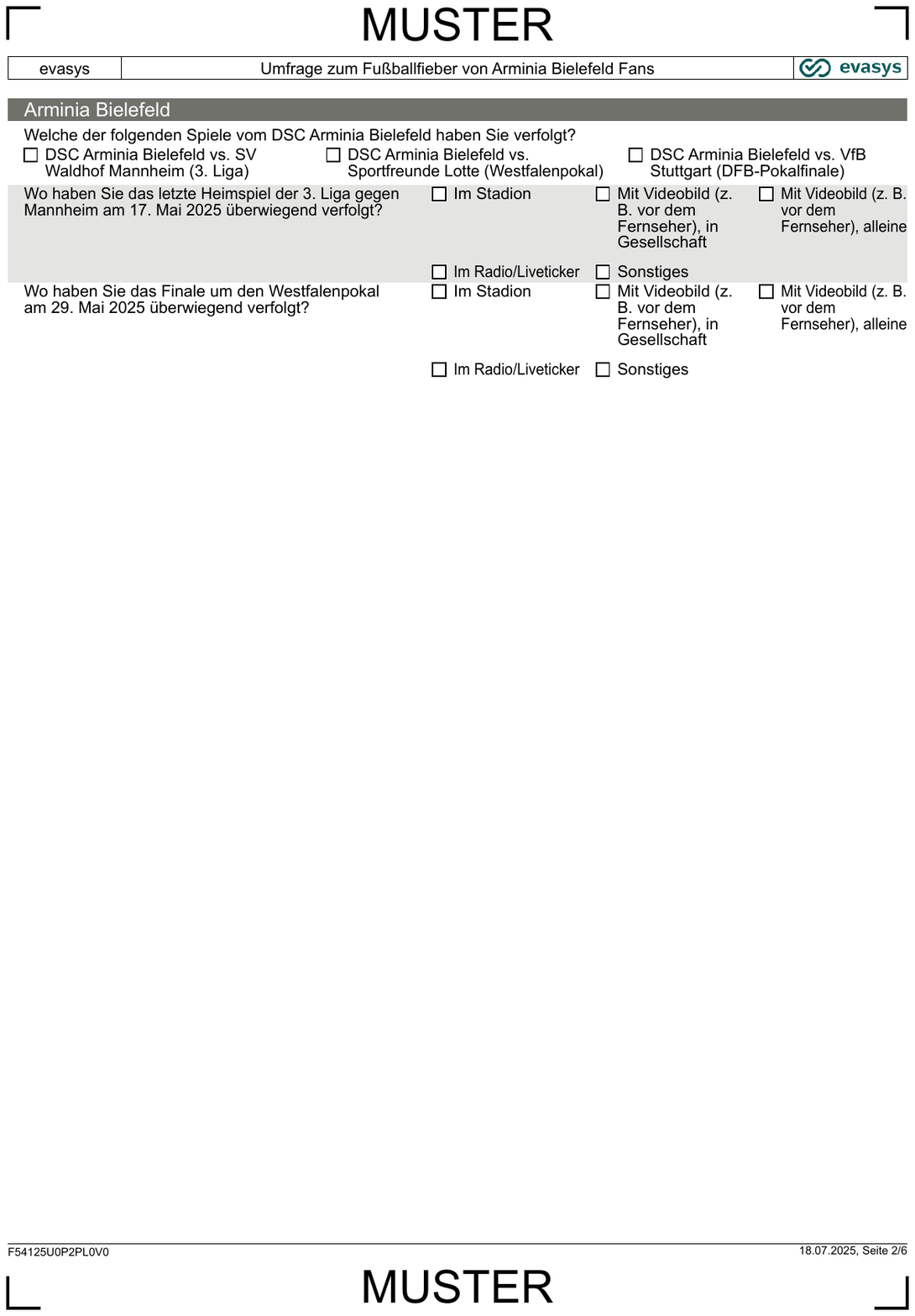}
\end{figure}

\begin{figure}
    \centering
    \includegraphics[width=\linewidth]{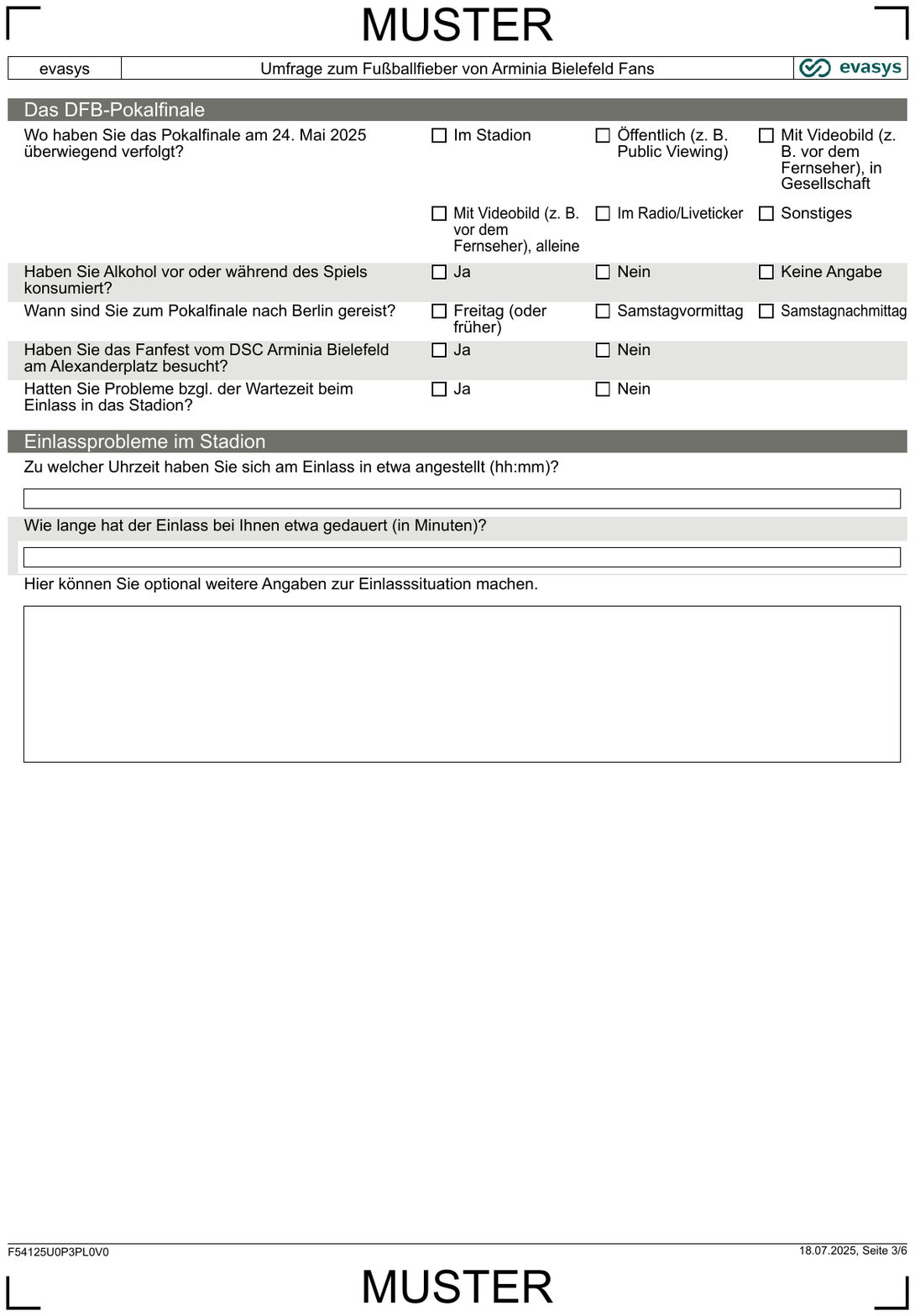}
\end{figure}

\begin{figure}
    \centering
    \includegraphics[width=\linewidth]{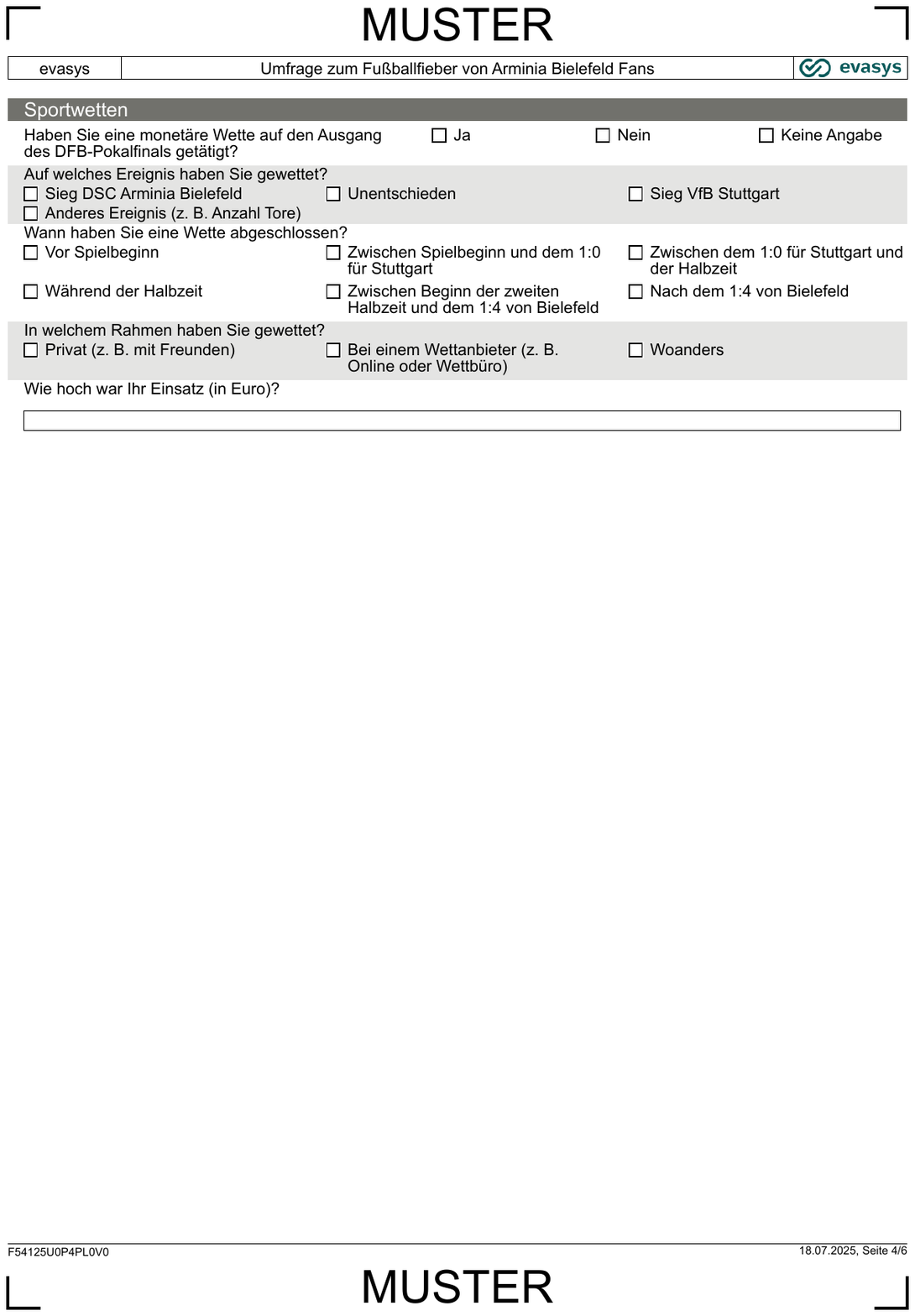}
\end{figure}

\begin{figure}
    \centering
    \includegraphics[width=\linewidth]{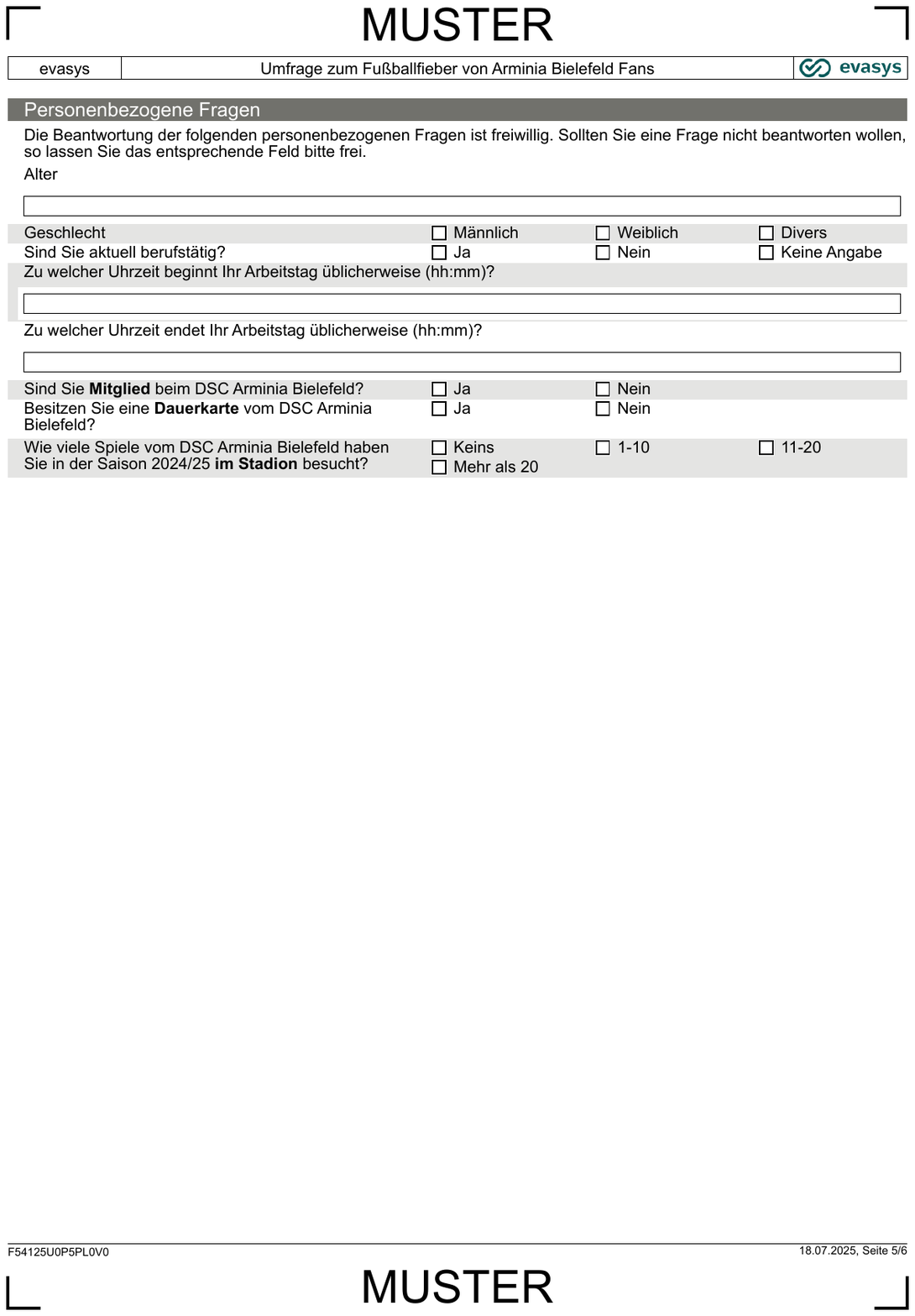}
\end{figure}

\begin{figure}
    \centering
    \includegraphics[width=\linewidth]{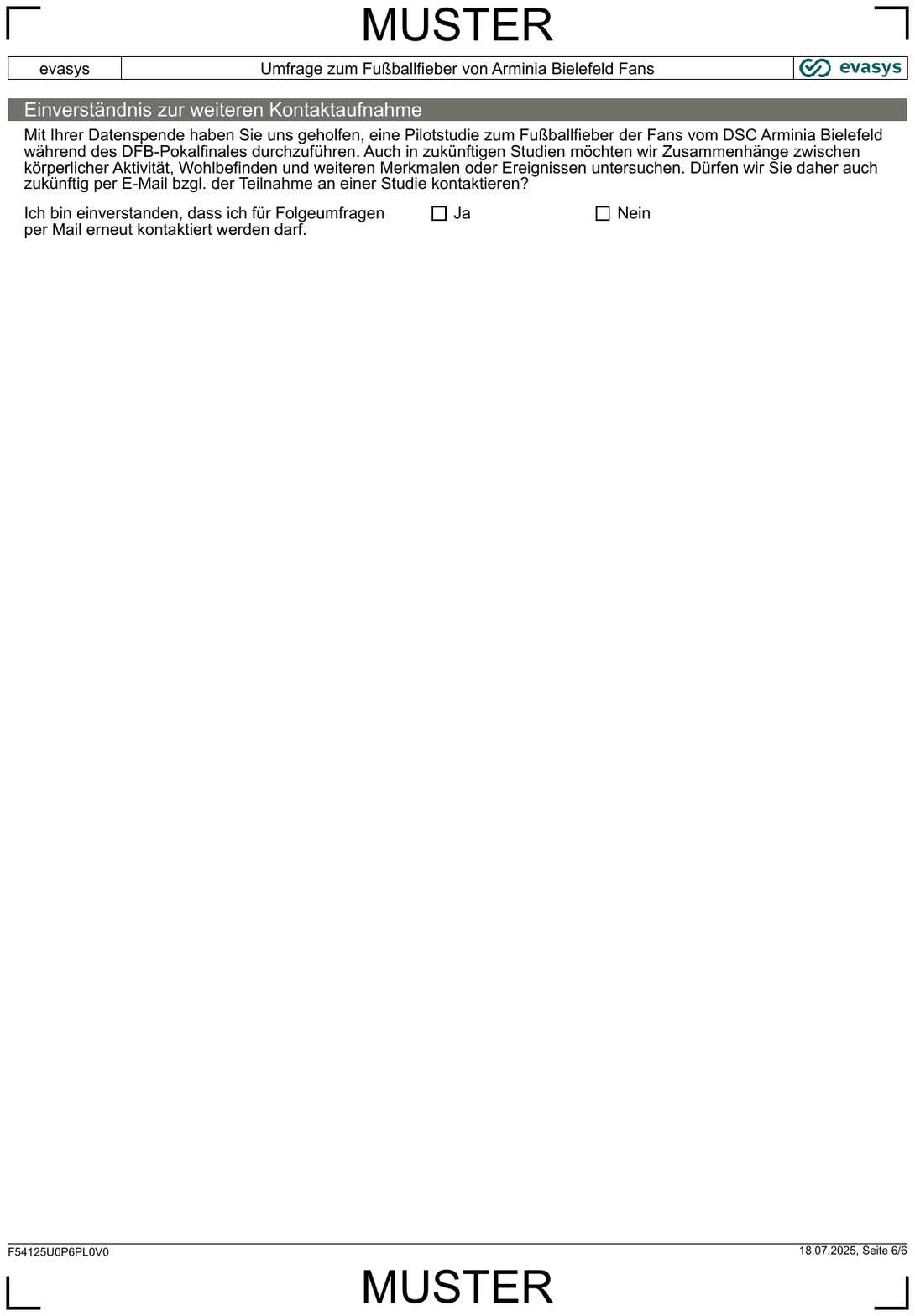}
\end{figure}

\end{appendices}

\end{spacing}
\end{document}